\documentclass[twocolumn,showpacs,preprintnumbers,amsmath,amssymb]{revtex4}

\usepackage{graphicx}
\usepackage{rotating}
\usepackage{dcolumn}
\usepackage{bm}

\begin{document}


\title{Observation of Feshbach-like resonances in collisions between ultracold molecules} 

\author{C. Chin,$^{1}$ T. Kraemer,$^{1}$ M. Mark,$^{1}$ J. Herbig,$^{1}$ P. Waldburger,$^{1}$ H.-C. N\"{a}gerl,$^{1}$ R. Grimm$^{1,2}$}
\affiliation{ $^1$Institut f\"{u}r Experimentalphysik,
Universit\"{a}t Innsbruck, Technikerstra\ss e 25, 6020 Innsbruck,
Austria \\
$^2$Institut f\"{u}r Quantenoptik und
Quanteninformation, \"{O}sterreichische Akademie der
Wissenschaften, 6020 Innsbruck, Austria}%

\date{\today}

\begin{abstract}
We observe magnetically tuned collision resonances for ultracold
Cs$_2$ molecules stored in a CO$_2$-laser trap. By magnetically
levitating the molecules against gravity, we precisely measure
their magnetic moment. We find an avoided level crossing which
allows us to transfer the molecules into another state. In the new
state, two Feshbach-like collision resonances show up as strong
inelastic loss features. We interpret these resonances as being
induced by Cs$_4$ bound states near the molecular scattering
continuum. The tunability of the interactions between molecules
opens up novel applications such as controlled chemical reactions
and synthesis of ultracold complex molecules.
\end{abstract}


\maketitle

The synthesis of ultracold molecules from ultracold atoms has
opened up new possibilities for studies on molecular matter-waves
\cite{csmol, namol, mbec}, strongly interacting superfluids
\cite{xoverexp}, high-precision molecular spectroscopy
\cite{boundbound} and coherent molecular optics \cite{moloptics}.
In all these experiments, control of the interatomic interaction
by magnetic fields plays an essential role in the association
process. When a two-atom bound state is magnetically tuned near
the quantum state of two scattering atoms, coupling from the
atomic to the molecular state can be resonantly enhanced. This is
commonly referred to as a Feshbach resonance \cite{feshbach}.

The success in controlling the interaction of ultracold atoms
raises the question whether a similar level of control can be
achieved for ultracold molecules. Resonant interactions between
molecules may lead to synthesis of complex objects beyond atomic
dimers. Furthermore, scattering processes for molecules involve
many novel reactive channels in comparison to the atomic
counterpart, e.g. collision induced dissociation, rearrangement or
displacement chemical reactions. Magnetic tunability of the
molecular interactions, similar to that resulting from atomic
Feshbach resonances, will lead to exciting perspectives to
investigate these chemical processes in regimes where quantum
statistics and quantum coherence play an important role.

In this Letter, we report the observation of magnetically tuned
collision resonances in an ultracold gas of Cs$_2$ molecules. The
ultracold dimers are created from an atomic Bose-Einstein
condensate (BEC) by use of a Feshbach ramp \cite{csmol} and
trapped in a CO$_2$-laser trap. We precisely measure the magnetic
moment of the molecules and observe an avoided crossing
\cite{rbmol} which allows us to transfer the molecules into
another state. In the new state, we discover two narrow inelastic
collision resonances. The resonance structure suggests that bound
states of two cesium molecules, or equivalently Cs$_4$ states,
induce the resonant scattering of molecules. These resonances,
which we interpret as Feshbach resonances for ultracold molecules,
may open the door to the synthesis of more complex molecules and
to the control of their chemical reactions.

\begin{figure}
\includegraphics[width=3in]{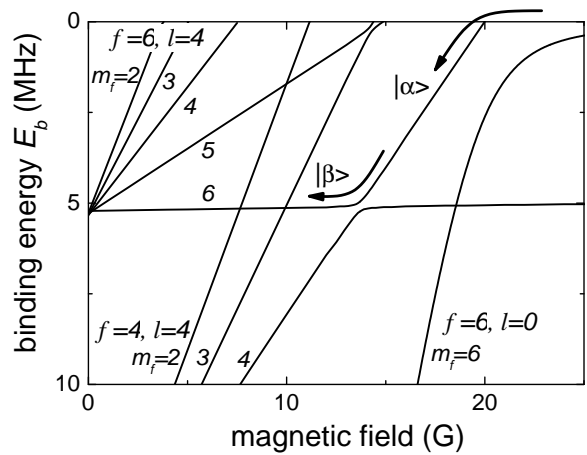}
\caption{Molecular energy structure below the scattering continuum
of two cesium atoms in the $|F=3, m_F=3\rangle$ state. The energy
of the dissociation threshold corresponds to $E_b=0$. The arrows
mark the paths to the molecular states we explore, which include
the creation of the molecules in $|\alpha\rangle$ via the atomic
Feshbach resonance at $19.84$~G \cite{csmol, cscreation} and an
avoided crossing to $|\beta\rangle$ at $\sim$13.6~G. Included are
only molecular states which can couple to the continuum via
Feshbach couplings up to $g-$wave interaction $(l\leq 4$,
$m_f+m_l=6$ and $m_f\geq2)$.} \label{fig1}
\end{figure}

The relevant molecular energy structure shown in Fig.~\ref{fig1}
is based on calculations done at NIST \cite{nist, csfesh}. The
dissociation threshold, providing the energy reference $E_b=0$, is
associated with two Cs atoms in the lowest ground state sublevel
$|F=3, m_{F}=3\rangle$, where $F$ and $m_F$ are the quantum number
of the atomic angular momentum and its projection, respectively.
As a result of the strong indirect spin-spin interaction of Cs
atoms \cite{spinspin}, coupling to molecular states with large
orbital angular momentum $l=4$ \cite{csfesh, chi03} leads to the
complexity of the energy structure shown in Fig.~\ref{fig1}. This
type of coupling is generally referred to as $g$-wave Feshbach
coupling.

We create the molecules in the bound state
$|\alpha\rangle\equiv|f=4,m_f=4;l=4, m_l=2\rangle$ via $g$-wave
Feshbach coupling at 19.84G \cite{csmol}, see Fig.~\ref{fig1}.
Here, $f$ is the internal angular momentum of the molecule, and
$m_f$ and $m_l$ are the projections of $f$ and $l$, respectively.
The molecular state $|\alpha\rangle$ is stable against spontaneous
dissociation for magnetic fields below 19.84~G and acquires larger
binding energies at lower magnetic fields. This is due to the
small magnetic momentum of $\sim0.95$~$\mu_B$ of this state as
compared to the atomic scattering continuum with
$\sim1.5$~$\mu_B$. At about 14~G, an avoided crossing to another
state $|\beta\rangle\equiv|f=6,m_f=6;l=4, m_l=0\rangle$ is induced
by the indirect spin-spin coupling. In this work, we ramp the
magnetic field adiabatically and explore the upper branch of the
avoided crossing.

Our experiment starts with an essentially pure atomic BEC with up
to $2.2\times 10^5$ atoms in a crossed dipole trap formed by two
CO$_2$ laser beams \cite{csreloaded, transfer}. We apply a
magnetic field of 20~G, slightly above the Feshbach resonance, and
a magnetic field gradient of 31.2~G/cm to levitate the atoms
\cite{csbec}. The CO$_2$-laser trap is roughly spherically
symmetric with a trapping frequency of $\omega\approx 2\pi\times$
20~Hz and a trap depth of 7~$\mu$K. The atomic density is $6\times
10^{13}$cm$^{-3}$ and the chemical potential is $k_B\times 20$~nK,
where $k_B$ is Boltzmann's constant.

To create the molecules, we first ramp the magnetic field from
$20.0$~G to $19.5$~G in 8~ms and then quickly change the field to
17~G to decouple the molecules from the atoms. Simultaneously, we
ramp the magnetic field gradient from 31.2 up to 50.2~G/cm. The
latter field gradient levitates the molecules \cite{csmol} and
removes all the atoms from the trap in 3~ms. As a consequence, we
obtain a pure molecular sample in the CO$_2$-laser trap with
typically $10^4$ molecules. The magnetic field ramping process
also leads to a small momentum kick on the molecules, which start
oscillating in the trap. After $\sim$100~ms, the oscillations are
damped out and the sample comes to a new equilibrium at a
temperature of $250$~nK with a peak density of $5\times
10^{10}$~cm$^{-3}$ and a phase space density of $10^{-2}$ to
$10^{-3}$. To measure the molecule number, we dissociate the
molecules into free atoms by reversely ramping the magnetic field
back above the resonance to 21~G. We then image the resulting atom
number \cite{csmol}.

A key parameter for a perfect levitation of the Cs$_2$ molecules
is the precise value of their magnetic moment \cite{csmol}. The
levitation field is crucial because the gravitational force is
much stronger than the trapping force of the CO$_2$ lasers. In
contrast to ground state atoms with only slow-varying magnetic
moment, the magnetic moment of the molecules can sensitively
depend on magnetic field as a result of the complex interactions
between molecular states, see Fig.~\ref{fig1}. Therefore, the
prerequisite to perform Cs$_2$ molecule experiments at different
magnetic fields is the knowledge of the molecular magnetic moment
for an accurate setting of the levitation field.

\begin{figure}
\includegraphics[width=3in]{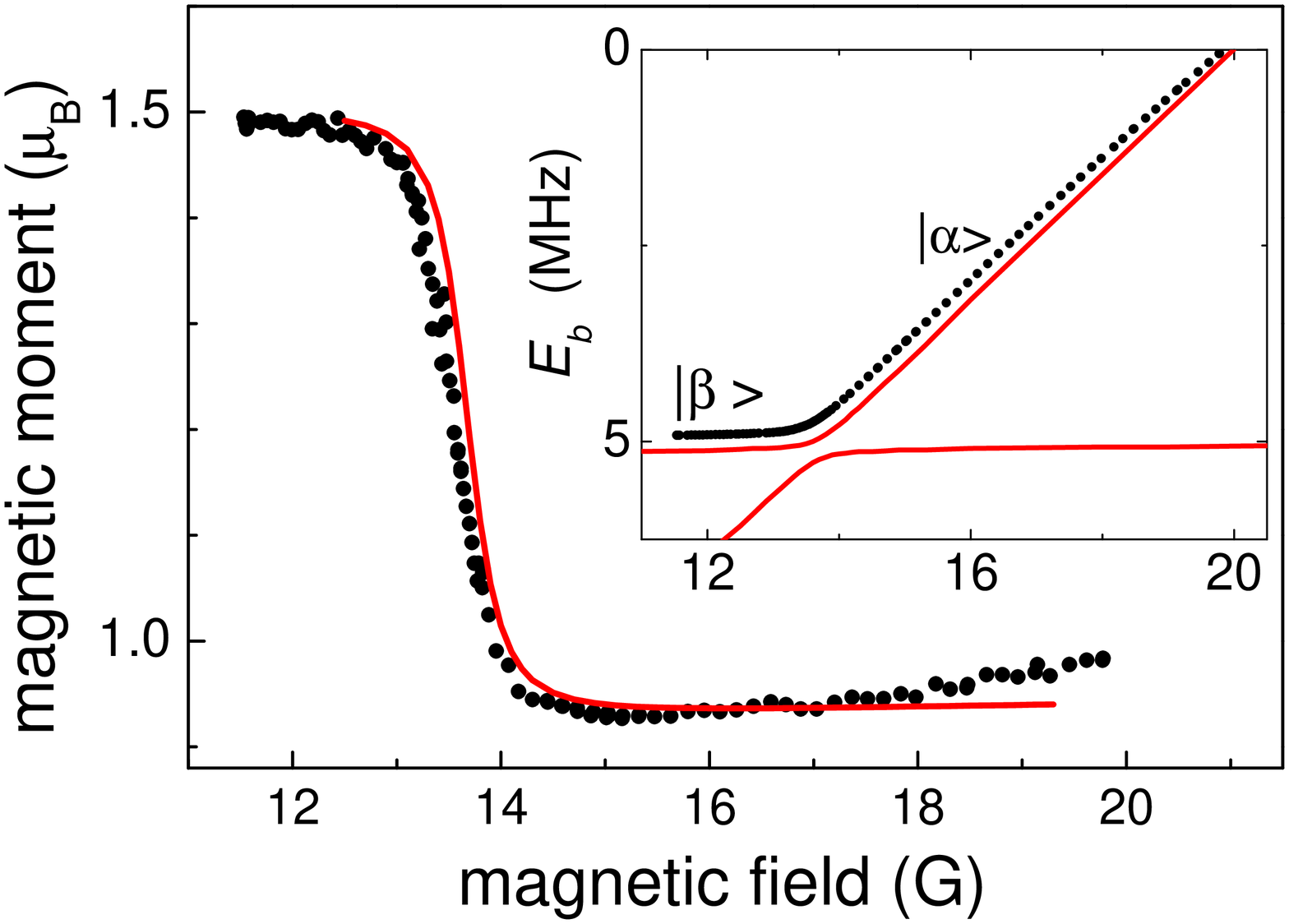}
\caption{Magnetic moment of the Cs$_2$ molecules. The measured
magnetic moment (solid circles) is compared to the NIST
calculation (solid line). The fast change at $\sim$13.6~G is
associated with an avoided crossing. In the inset, we derive the
molecular binding energy (solid circles) by integrating the
measured magnetic moment. Binding energies from the NIST
calculation (solid lines) for both branches of the avoided
crossing between state $|\alpha\rangle$ and state $|\beta\rangle$
are shown, see also Fig.~\ref{fig1}.} \label{fig2}
\end{figure}

We map out the magnetic moment of the molecules over the range of
11.5~G to 19.8~G. This is realized by a two-step process: First,
we slowly tune the magnetic field in 60~ms  to a desired value and
find a corresponding magnetic field gradient which can
approximately keep the molecules near the center of the
CO$_2$-laser trap. Second, after a hold time of 500~ms needed for
the ensemble to come to an equilibrium, we measure the position of
the cloud. The location of the molecular cloud provides a very
sensitive probe to the residual imbalance of the magnetic force
and gravity. Given a small vertical displacement of the molecules
relative to the trap center $\delta z$ for a local magnetic field
$B$ and a field gradient $B'$, the magnetic moment is then
$\mu(B)=(2m\omega^2\delta z+2mg)/B'$. Here $2m$ is the molecular
mass, and $g$ is the gravitational acceleration. Independent
measurements based on releasing the molecules into free space
\cite{csmol} confirm the accuracy of the above method to
$0.01$~$\mu_B$.

The measured magnetic moments of the molecules show the expected
behavior in the range of $11.5$~G to $19.8$~G, see
Fig.~\ref{fig2}. We find that the magnetic moment slowly decreases
from 0.98~$\mu_B$ to 0.93~$\mu_B$ as the magnetic field is lowered
from 19.8~G. For magnetic fields below $\sim14$~G, the magnetic
moment quickly rises and levels off at 1.5~$\mu_B$. This behavior
is readily explained by the avoided crossing at 13.6~G
(Fig.~\ref{fig1} and \ref{fig2}), which transfers the molecules
from state $|\alpha\rangle$ with $\mu\approx0.9$~$\mu_B$ to
$|\beta\rangle$ with $\mu\approx1.5$~$\mu_B$. Below 11.5~G, a new
avoided crossing to a very weakly coupled $l=8$ molecular state
occurs \cite{l8}. We observe fast loss of the molecules since our
current apparatus cannot produce a sufficient levitation field to
support the molecules against gravity in this new state.

Our measurement agrees excellently with the NIST calculation
\cite{nist, csfesh} within the 200~mG uncertainty from the
multi-channel calculation, see Fig.~\ref{fig2}. We evaluate the
molecular binding energy based on integrating the measured
magnetic moments. Here the integration constant is fixed by the
fact that the molecular binding energy is zero at the atomic
Feshbach resonance $B=19.84$~G. The result shown in the inset of
Fig.~\ref{fig2} gives very good agreement with the theoretical
calculation within the energy uncertainty of 0.25~MHz \cite{nist}.
By fitting our binding energies to a simple avoided crossing
model, we determine the crossing to be $B_{\rm{cross}}=13.55(4)$~G
and the coupling strength, half the energy splitting between the
two eigenstates at $B_{\rm{cross}}$, to be $h\times$150(10)~kHz.
Here $h$ is Planck's constant.

\begin{figure}
\includegraphics[width=3.2in]{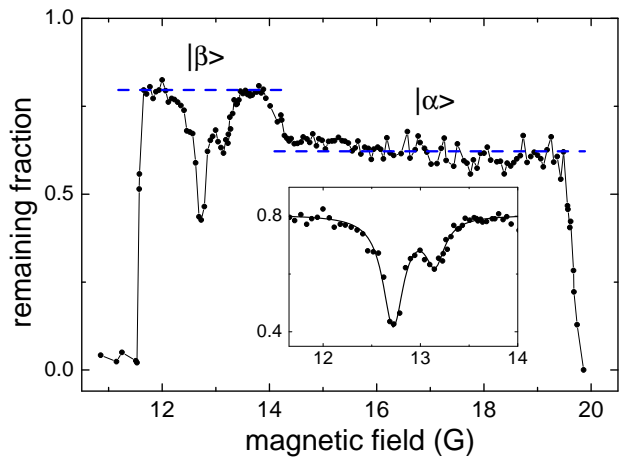}
\caption{Remaining fraction of optically trapped molecules after a
storage time of 300~ms. Initially, there are 11000 molecules at a
peak density of $6\times10^{10}$cm$^{-3}$ and a temperature of
$250$~nK. The dashed lines mark the background loss rates in state
$|\alpha\rangle$ and in state $|\beta\rangle$. The two loss
resonances for $|\beta\rangle$ are fit by a sum of two Lorentzian
profiles (inset).} \label{fig3}
\end{figure}

To investigate the interactions between molecules, we measure the
inelastic collision loss after a trapping time of 300~ms
(Fig.~\ref{fig3}). For molecules in state $|\alpha\rangle$
(14~G$<B<$19.8~G), the fractional loss is about $\sim 40\%$. In
this molecular state, we do not see any strong magnetic field
dependence. When the magnetic field is tuned near the Feshbach
resonance at 19.8~G, molecules dissociate into free atoms, which
leave the trap.

In state $|\beta\rangle$ ($11.5$~G$<B<$13.6~G), the behavior of
the molecules is strikingly different. We observe a weaker
background loss of $\sim 20\%$ and two pronounced resonances with
a fractional loss of up to $60\%$. An expanded view in the inset
of Fig.~\ref{fig3} shows that the ``double peak" structure can be
well fit by a sum of two Lorentzian profiles. From the fit, we
determine the resonance positions to be $12.72(1)$~G and
$13.15(2)$~G with full widths of $0.25$~G and $0.24$~G,
respectively. Note that due to the levitation gradient field, the
inhomogeneity across the molecular sample is as large as $0.15$~G
in state $|\beta\rangle$, which suggests that the intrinsic widths
of these resonances are less than the observed values.

The observed resonances cannot be explained by single-molecule
effects based on the Cs$_2$ energy structure, which is precisely
known to very high partial waves \cite{csfesh, nist}. Beyond
single-molecule effects, the observed resonance structure strongly
suggests that bound states of Cs$_2$ molecules (Cs$_4$ states) are
tuned in resonance with the scattering state of the molecules and
induce Feshbach-like couplings to inelastic decay channels. For
Cs$_2$ molecules, the appearance of Cs$_4$ bound states near the
scattering continuum is not surprising considering the complexity
of interaction between Cs atoms and the additional rotational and
vibrational degrees of freedom.

\begin{figure}
\includegraphics[width=3in]{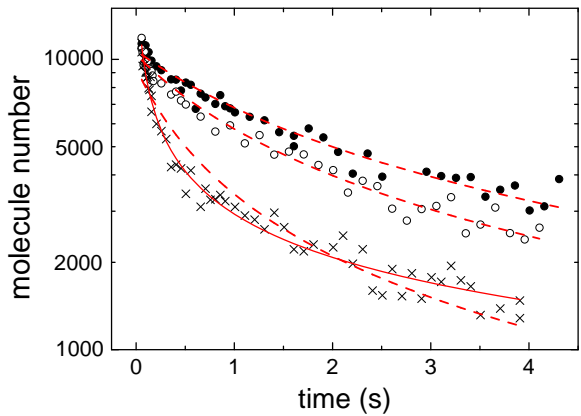}
\caption{Time evolution of the molecule number in the CO$_2$-laser
trap for molecules in state $|\alpha\rangle$ at 15.4G (open
circles), in state $|\beta\rangle$ at 12.1~G (off-resonance, solid
circles) and at 12.7~G (on resonance, crosses). Fits based on
two-body loss (dashed lines) work well for 15.4~G, and 12.1~G. A
fit based on three-body loss (solid line) works better for
12.7~G.} \label{fig4}
\end{figure}

To verify that the loss is indeed due to collisions between
molecules, we measure the time evolution of the molecular
population in the CO$_2$-laser trap. Starting with 11000 molecules
prepared at different magnetic fields, we record the molecule
number after various wait times, as shown in Fig.~\ref{fig4}.
Three magnetic field values are chosen here: $15.4$~G where the
molecules are in state $|\alpha\rangle$, $12.1$~G where the
molecules are in state $|\beta\rangle$ and are away from the
resonance, and $12.7$~G where the molecules are on the strong
molecular resonance, see Fig.~\ref{fig4}.

The number of trapped molecules shows a non-exponential decay,
which provides a clear signature of density-dependent collision
processes. To determine the underlying collision processes, we
model the loss based on a two-body or a three-body loss equation.
Assuming a Gaussian distribution for the thermal ensemble in a
harmonic trap with a constant temperature and that the collision
loss rate is slow compared to the thermalization rate, we fit the
measured molecule numbers to the two- and three-body decay
equation, see in Fig.~\ref{fig4}.


For $15.4$~G and $12.1$~G, we find that the two-body equation
provides excellent fits. The two-body coefficients are
$5\times10^{-11}$cm$^3$/s at 15.4~G and $3\times10^{-11}$cm$^3$/s
at 12.1~G. We, however, cannot rule out the possibility that
three-body processes also play a role. The measured collision rate
coefficients are similar to the measurements from the MIT group on
Na$_2$ \cite{nadisso}, and are an order of magnitude below the
unitarity limit of $2 h/mk=4\times 10^{-10}$cm$^3$/s, where $k$ is
the characteristic collision wave number associated with the
temperature of the sample.

At 12.7~G, where the molecules are on the strong resonance, we
find that the three-body equation actually provides a better fit
than the two-body fit with a three-body loss coefficient at
$6\times 10^{-20}$cm$^6$/s, see Fig.~\ref{fig4}. This value,
however, is much too high compared to the three-body unitarity
limit of $96\pi h/mk^4=2\times 10^{-23}$cm$^6$/s \cite{sun03}. One
alternative explanation is that on resonance, the fast collision
loss rate might leave the molecules insufficient time to reach
thermal equilibrium. By fitting the resonance data in the first
200~ms with the two-body loss model, we determine the two-body
loss coefficient to be $2\times10^{-10}$cm$^3$/s, which indeed
approaches the unitarity limit of $4\times10^{-10}$cm$^3$/s.

The roles of two-body and three-body processes near a collision
resonance have been extensively studied for ultracold atoms. For
two-body losses, colliding atoms can only decay into two free
atoms with lower internal energies. For three-body losses, the
outgoing channels, in general, are either three free atoms or one
atom and one dimer. For molecules, however, many more outgoing
channels are possible. For instance, two resonantly interacting
Cs$_2$ molecules can decay into four Cs atoms, two Cs atoms and
one Cs$_2$ dimer, or one Cs atom and one Cs$_3$ trimer. In the
case of three-molecule collisions, there exist, in principle, ten
different types of decay processes.

More intriguingly, the observed magnetically tuned Feshbach-like
resonances bring in fascinating prospects for a controlled
synthesis of Cs$_4$ tetramers in a single four-body bound state.
This is analogous to the formation of the Cs$_2$ dimers near the
atomic Feshbach resonances. The tunability of the interactions in
molecular quantum gases can potentially open up the door to
ultracold chemistry and to few-body physics beyond simple atoms
and diatomic molecules.

We greatly thank E. Tiesinga and P. S. Julienne for stimulating
discussions, and in particular, for providing us with the
theoretical calculation on the Cs$_2$ energy structure. We
acknowledge support by the Austrian Science Fund (FWF) within SFB
15 (project part 16) and by the European Union in the frame of the
Cold Molecules TMR Network under Contract No.\ HPRN-CT-2002-00290.
M.M. is supported by DOC [Doktorandenprogramm der
\"{O}sterreichischen Akademie der Wissenschaften]. C.C.\ is a
Lise-Meitner research fellow of the FWF.


\begin{references}
\bibitem{transfer}
We adiabatically transfer the atomic BEC from a near-infrared
laser trap, where the atomic BEC is formed \cite{csreloaded}, to
the CO$_2$-laser trap by simultaneously reducing the near-infrared
laser power to zero and ramping the CO$_2$ lasers to full power in
1~s.
\bibitem{l8}
This $l=8$ state, not shown in Fig.~\ref{fig1}, is very weakly
coupled to the state $|\beta\rangle$. Investigations on this state
are currently in progress.
\end{references}
\end{document}